\documentclass[%
 reprint,
 amsmath,amssymb,
 aps,
]{revtex4-1}

\usepackage{graphicx}
\usepackage{dcolumn}
\usepackage{bm}

\begin{document}


\title{Numerical solution of Mathisson-Papapetrou-Dixon equations\\
for spinning test particles in a Kerr metric}
\thanks{One of the authors N.V. is grateful to the Pontificia Universidad Javeriana in Bogot\'a for financial support and a doctoral scholarship, and to Professors Roman Plyatsko and J.~Alfonso Leyva for their helpful suggestions}%

\author{Nelson Velandia}
 \altaffiliation[Also at ]{Physics Department, Pontificia Universidad Javeriana. navelandia@javeriana.edu.co, navelandiah@unal.edu.co}
\author{Juan Manuel Tejeiro}%
 \email{jmtejeiros@unal.edu.co}
\affiliation{%
 Universidad Nacional de Colombia\\
 Pontificia Universidad Javeriana 
}%


\date{\today}

\begin{abstract}
In this work we calculate some estimations of the gravitomagnetic clock effect,
taking into consideration not only the rotating gravitational field of the
central mass, but also the spin of the test particle, obtaining values for $\Delta t=t_{+}-t_{-}=2.5212079035\times10^{-8} \textrm{s}$. We use the formulation of Mathisson-Papapetrou-Dixon equations for this problem in a Kerr metric. In order to
compare our numerical results with previous works, we consider initially only the
equatorial plane and also apply the Mathisson-Pirani supplementary spin
condition for the spinning test particle. 
\begin{description}

\item[PACS numbers]
98.80Jk
 
\end{description}
\end{abstract}

\maketitle


\section{Introduction}

In the last decades, important advances have been made in the study of the
gravitomagnetic clock effect. Beginning with the seminal work by Cohen and
Mashhoon \cite{cohen 1}. In which they presented the influence of the
gravitomagnetic field to the proper time of an arbitrary clock about a rotating
massive body. In their paper, Cohen and Mashhoon, also showed the possibility
of measuring this effect. In this work, we present a theoretical value for the
gravitomagnetic clock effect of a spinning test particle orbiting around a
rotating massive body.

According with the literature, we find different complementary ways that
study the phenomena in regard to the gravitomagnetism clock effect. The
first way take two family of observers. The first is the family of static
observer (or threading observers) with four-velocity $m=M^{-1}\partial _{t}$
and world lines along the time coordinate lines. The second famili is the
ZAMO\'{}
s (or slicing observers) with four-velocity $n=N^{-1}\left( \partial
_{t}-N^{\phi }\partial _{\phi }\right) $ and world lines orthogonal to the
time coordinate hypersurfaces \cite{bini 2002}, \cite{bini feb}, \cite{bini 2001}. They obtain, in the threading
point of view, the local spatial angular direction as
\begin{equation}
    dt = M_{\phi}\rightarrow \frac{dt}{d\phi} = M_{\phi} = \zeta_{\left(\text{th}\right)}^{-1}
\end{equation}

which gives an inverse angular velocity or the change with respect to angle
of the time coordinate. Since is angular velocity, Bini \textit{et al.}
integrate the coordinate time for one complete revolution both in a
direction and in opposite direction \cite{bini 2002}. Then the physical components of the
velocities are related to the coordinate angular velocity.

This group study the case when the particle has spin. They take the
Frenet-Serret frame (FS) associated to worldline of the test particle and
calculates with help of the angular velocity the evolution equation of the
spin tensor in terms of the FS intrinsic frame \cite{bini 2004 jul}, \cite{bini 2004 oct}. The work of this group considers the MPD equations
and their su-pplementary conditions for the spin and give their answer in
terms of angular velocity.

The second group integrates around a closed contour. They take the time for
this loop when the test particle rotates in clockwise and the test particle
in opposite sense \cite{cohen 1}, \cite{cohen moses}.

A third group deduces the radial geodesic equation from the line element in
the exterior field of a rotating black hole. With this equation yields the
solution and calculate the inverse of the azimuthal component of four
velocity. Then they introduce the first order correction to the angular
velocity 
\begin{equation}
\Omega \equiv \left( \frac{d\varphi }{dt}\right) _{0}\left(
1+S\Omega ^{\ast }\right),    
\end{equation}
and obtain the differentiate between
prograde and retrograde orbits and integrate from zero to 2$\pi $. The clock
effect is the difference of theses two orbits \cite{faruque 2004}, \cite{faruque 2006}, \cite{shahrear 1}.

The fourth group takes some elements of electromagnetism and does an analogy
between Maxwell equations and Einstein linealized equations \cite{iorio 2005}, \cite{iorio 2001}, \cite{iorio lich}, \cite{iorioL 1}, \cite{lichte 1}, \cite{lich 2003}, \cite{mash 2001}, \cite{mash 1999}, \cite{mash 1998}. Finally the group that makes a geometric
treatment of the gravitomagnetic clock Effect \cite{tartaglia 1}, \cite{tartaglia 2000}.

According with other papers that work the MPD equations, the novelty of our
work is that we calculate numerically the full set of MPD equations for the
case of a spinning test particle in a Kerr metric. Secondly, we take the
spin without restrictions in its velocity and spin orientation. In the paper
by Kyrian and Semer\'{a}k the third example is refered to the particular
case when the spin is orthogonal to the equatorial plane in a Kerr metric \cite{kyrian 1}.

In this paper, our aim, it is not only describing the
trajectories of spinning test particles, but also to study the clock effect.
Therefore, we calculate numerically the trajectory both in a sense and in
the other for a circular orbit. We measure the delay time for three
situations: two spinless test particles are traveling in the same circular
orbit, two spinning test particles with its spin value orthogonal to
equatorial plane and two spinning test particles without restrictions in its
spin orientation.

In this paper, we use the set of Mathisson-Papapetrou-Dixon equations (MPD)
presented by Plyatsko, R. \textit{et al.} \cite{plyatsko 1} and extend this
approach, considering the spin of the particle without restrictions in its
orientation, while Plyatsko \textit{et al.} only take a constant value for the
spin in its magnitude and orientation; we allow both to be varied, using the
full set of MPD equations and the Mathisson-Pirani (MP) supplementary spin condition.

In the literature, one can find different conditions to fix the center of
mass, leading to different kinematical behaviours of the test particles. One of
the features of the MPD equations is the freedom to define the
representating worldline to which $u^{\alpha}$ is tangente vector. Therefore the worldline can be
determined from physical conside-rations. The first condition is presented by
Corinaldesi and Papapetrou (CP, 1951) which is given by
\begin{equation}
S^{i 0}=0 ,
\end{equation}
which depends on coordinates. For this condition, the worldline is
straight and its tangent $u^{\alpha}$ is parallel to the four momentum
$p^{\mu}$ \cite{kyrian 1}. The second condition is the Mathisson-Pirani
condition (MP, 1956):
\begin{equation}
u_{\alpha}S^{\alpha\beta}=0
\label{mp}
\end{equation}
in this condition the reference worldline is the center of mass as measured in the rest frame of the observer of velocity $u^{\alpha}$ \cite{mathisson 1},
\cite{pirani 1}. This condition does not fix a unique worldline and $u^{\alpha}$ is uniquely defined by $p^{\alpha}$ and $S^{\alpha\beta}$. If one uses this condition, the
trajectory of the spinning test particle is represented by helical motions.
Costa \textit{et al. }\cite{costaF 1} explain that these motions are
physically possible. We use this condition when working with the MPD equations in the
case of a spinning test particle orbiting a rotating massive body.

The third condition is introduced by Tulczyjew and Dixon (TD, 1959) and written which is given by
\begin{equation}
p_{\alpha}S^{\alpha\beta}=0
\end{equation}
where
\begin{equation}
p^{\sigma}=mu^{\sigma}+u_{\lambda}\frac{DS^{\sigma\lambda}}{ds}
\end{equation}
is the four momentum. This condition implies that the worldline is straight and its tangent $u^{\alpha}$ is
parallel to $p^{\alpha}$ and the spin is constant \cite{dixon 1}. This
condition is cova-riant and guarantees the existence and uniqueness of the
respective worldline \cite{beiglbock 1}.

The fourth is given by Newton and Wigner (NW) which is a combination
of the TD and MP conditions
\begin{equation}
S^{\alpha\beta}\zeta_{\alpha}=0
\end{equation}
with $\zeta_{\alpha}:=p_{\alpha}+Mu_{\alpha}$ and $u_{\alpha}$ being a
timelike vector. This condition provides an implicit relation between the
four-momentum and the wordline\'{}s tangent vector.

The fifth condition is called Ohashi-Kyrian-Semerak (OKS, 2003):
\begin{equation}
w_{\alpha}S^{\alpha\beta}=0
\end{equation}
where $w^{\alpha}$ is some time-like four vector which parallely transports along the representative worldline \cite{kyrian 1}:
\begin{equation}
w_{\alpha}w^{\alpha}=-1, \ \dot{w}
^{\alpha}=0.   
\end{equation}
For the study of spinning test particles, we use the equations of motion for a
spinning test particle in a gravitational field without any restrictions to
its velocity and spin orientation \cite{plyatsko 1}. In this paper,
we use the MPD equations presented by Plyatsko, R. \textit{et al.
}\cite{plyatsko 1}. They\textit{\ }yield the full set of
Mathisson-Papapetrou-Dixon equations (MPD equations) for spinning test
particles in the Kerr gravitational field \cite{plyatsko 1}, where
they integrate nume-rically the MPD equations for the particular case of the
Schwarzschild metric.

For the scope of this work, we will take the MPD equations of motion for a
Kerr metric, and additionally we will include the spin of the test particle.
This calculation has been made with the Mathisson-Pirani supplementary
condition; the trajectories have been obtained by numerical integration, using
the Runge-Kutta algorithm \cite{press 1}.

Presently, there exists an interest in the study of the effects of the spin on
the trajectory of test particles in rotating gravitational fields \cite{wen-biao 1}. The importance of this topic increases when dealing with
phenomena of astrophysics such as accretion discs in rotating black holes, gravitomagnetics effects \cite{faruque 2004} or gravitational waves induced by spinning particles
orbiting a rotating black hole \cite{tanaka 1996}, \cite{mino 1}.

The motion of particles in a gravitational field is given by the geodesic
equation. The solution to this equation depends on the particular conditions of
the problem, such as the rotation and spin of the test particle, among o-thers;
therefore there are different methods for its solution \cite{carter 1}, \cite{abramowicz 1}.

Basically, we take two cases in motion of test particles in a gravitational
field of a rotating massive body. The first case describes the trajectory of a
spinless test particle, and the second one the trajectory of a spinning test
particle in a massive rotating body. In the case of the spinless test particles,
some authors yield the set of equations of motion for test particles orbiting
around a rotating massive body. The equations of motion are considered both
in the equatorial plane \cite{bardeen 1}, \cite{wilkins 1}, \cite{calvani 1}, and in the non-equatorial plane \cite{wilkins 1}, \cite{stog 1}, \cite{teo 1} (Kheng, L., Perng, S., and
Sze Jackson, T.: Massive Particle Orbits Around Kerr Black Holes.
Unpublished). For the study of test particles in a rotating field, some
authors have solved for particular cases the equations of motion both for
spinless and for spinning test particles of circular orbits in the equatorial
plane of a Kerr metric \cite{tartaglia 1}, \cite{tanaka 1996}, \cite{bardeen 1}, \cite{suzuki 1}, \cite{mash 1}, \cite{dadhich 1}, \cite{bini 2005}, \cite{tod 1}.

With the aim of proving the equations of motion with which we worked, solve numerically the set of equations of motion obtained via MPD equations both for the spinless particles and for spinning particles in the equatorial plane and will compare our results with works that involve astronomy, especially the study of spinning test particles around a rotating central source. We take the
same initial conditions in the two cases for describing the trajectory of both a
spinless particle and a spinning particle in the field of a rotating massive
body. Then, we compare the Cartesian coordinates ($x,y,z$) for the trajectory
of two particles that travel in the same orbit but in opposite directions.

For the numerical solution, we give the full set of MPD equations
explicitly, while that Kyrian and Semerak only name them. Also, we give the
complete numerical solution. Kyrian and Semer\'{a}k integrate with a step of 
$M/100=1\times 10^{-2}$ while we integrate with a step of $%
n=2^{-22}=2.384185\times 10^{-7}$. In the majority of cases, the solutions
are partial because it is impossible to solve analytically a set of eleven
coupled differential equations. The recurrent case that they solve is a
spinning test particle in the equatorial plane and its spin value is
constant in the time ($S_{\perp }=$constant).

This work is organized as follows. In Section 2 we give a brief introduction
to the MPD\ equations that work the set of equations of motion for test
particles, both spinless and spinning in a rotating gravitational field. From
the MPD equations, we yield the equations of motion for spinless and spinning
test particles. Also, we will give the set of the MPD equations given by Plyatsko
\textit{et al.}\cite{plyatsko 1} in schematic form to work the equations of motion in a Kerr metric. In Section 3 and 4, we present the
gravitomagnetic clock effect via the MPD equations for
spinless and spinning test particles. Then, in Section 5, we perform integration and the respective numerical
comparison of the coordinate time ($t$) for spinless and spinning test particles in the
equatorial plane. Finally we make a numerical comparison of the trajectory in Cartesian
coordinates for two particles that travel in the same orbit, but in opposite
directions. In the last section, conclusions and some future works. We shall use geometrized units; Greek indices run from 1 to 4 and Latin indices run from 1 to 3. The metric signature (-,-,-,+) is chosen.

\section{Introduction to the Mathisson-Papapetrou-Dixon equations}

In general the MPD equations \cite{mathisson 1}, \cite{dixon 1}, \cite{papapetrou 1} describe the dynamics of extended
bodies in the general theory of relativity which includes any gravitational
background. These
equations of motion for a spinning test particle are obtained in terms of an
expansion that depends on the derivatives of the metric and the multipole
moments of the energy-momentum tensor ($T^{\mu\nu}$) \cite{dixon 1}
which describes the motion of an extended body. In this work, we will take a
body small enough to be able to neglect higher multipoles. According to this
restriction the MPD equations are given by
\begin{equation}
\frac{D}{ds}\left(  mu^{\lambda}+u_{\lambda}\frac{DS^{\lambda\mu}}{ds}\right)
=-\frac{1}{2}u^{\pi}S^{\rho\sigma}R_{\pi\rho\sigma}^{\lambda},\label{mov1}
\end{equation}

\begin{equation}
\frac{D}{ds}S^{\mu\nu}+u^{\mu}u_{\sigma}\frac{DS^{\nu\sigma}}{ds}-u^{\nu
}u_{\sigma}\frac{DS^{\mu\sigma}}{ds}=0,\label{mov2}
\end{equation}
where the covariant derivative is given by $D/ds$ , the antisymmetric tensor
$S^{\mu\nu}$, $R^{\lambda}{}_{\pi\rho\sigma}$ is the curvature tensor, and
$u^{\mu}=dz^{\mu}/ds$ is the particle\'{}s four-velocity. We do not have the evolution equation for $u^{\mu}$ and
it is necessary to single out the center of mass which determines the world
line as a representing path and specifies a point about which the momentum and
spin of the particle are calculated. The worldline can be determined from
physical considerations \cite{karpov 1}. In general, two conditions are
typically imposed: The Mathisson-Pirani supplementary condition (MP) $u_{\sigma}S^{\mu\sigma}=0$  \cite{mathisson 1}, \cite{pirani 1} and the Tulczyjew-Dixon condition $p_{\sigma}S^{\mu\sigma}=0$ \cite{dixon 1}.
We found that if we contract the equation
\begin{equation}
\frac{DS^{\alpha\beta}}{ds}=2P^{[\alpha}u^{\beta]},
\end{equation}
with the four velocity $u^{\alpha}$, we obtain%
\begin{equation}
P^{\beta}=mu^{\beta}-u_{\alpha}\frac{DS^{\alpha\beta}}{ds},
\end{equation}
where $m\equiv-P^{\alpha}u_{\alpha}$. Given the MP condition, the four
momentum is not para-llel to its four velocity $u^{\beta}$; therefore, it is
said to possess ``hidden momentum". This last equation can be written as
\begin{equation}
P^{\beta}=mu^{\beta}+S^{\alpha\beta}a_{\alpha},
\end{equation}
where $a^{\alpha}=Du^{\alpha}/ds$ is the acceleration. This acceleration
results from an interchange between the momentum $mu^{\beta}$ and hidden
momentum $S^{\alpha\beta}a_{\alpha}$. These variations cancel out at every
instant, keeping the total momentum constant \cite{costa 1}. The above equation
can be expressed as
\begin{equation}
P^{\beta}=P_{kin}^{\beta}+P_{hid}^{\beta},
\end{equation}
where $P_{kin}^{\beta}=mu^{\beta}$ is the kinetic momentum associated with the
motion of the centroid and the component $P_{hid}^{\beta}=S^{\alpha\beta
}a_{\alpha}$ is the hidden momentum. In this case, if the observer were in the center of mass, he would see its centroid at rest then we would have a helical solution.

To obtain the set of MPD equations, we take the MP condition which has three independent relationships between $S^{\mu\sigma}$ and
$u_{\sigma}$. By this condition $S^{i4}$ is given by
\begin{equation}
S^{i4}=\frac{u_{k}}{u_{4}}S^{ki}
\end{equation}
with this expression we can use the independent components $S^{ik}$. Sometimes
for the representation of the spin value, it is more convenient to use the vector
spin, which in our case is given by 
\begin{equation}
S_{i}=\frac{1}{2u_{4}}\sqrt{-g}\epsilon_{ikl}S^{kl}   \end{equation}
where $\epsilon_{ikl}$ is the spatial L\'{e}vi-Civit\`{a} symbol.

When the space-time admits a Killing vector $\xi^{\upsilon
}$, there exists a property that includes the covariant derivative and the
spin tensor, which gives a constant and is given by \cite{dixon 1979}
\begin{equation}
p^{\nu}\xi_{\nu}+\frac{1}{2}\xi_{\nu,\mu}S^{\nu\mu}=\textrm{ constant,}
\label{16}
\end{equation}
where $p^{\nu}$ is the linear momentum, $\xi_{\nu,\mu}$ is the covariant
derivative of the Killing vector, and $S^{\nu\mu}$ is the spin tensor of the
particle. In the case of the Kerr metric, one has two Killing vectors, owing
to its stationary and axisymmetric nature. In consequence, Eq. (\ref{16})
yields two constants of motion: the total energy $E$\ and the component
$z$\ of the angular momentum $J$ \cite{iorio 1}.

\subsection{MPD equations for a spinning test particle in a metric of a rotating
body}

Given that the spinning test body is small enough compare with the
characte-ristic length, this body can be considered as a test particle. In this section,the equations of
motion (Eqs. \ref{mov1} and \ref{mov2}) of the test particle are firstly introduced in the case when the particle is orbiting in an axisymmetric and stationary spacetime. Then, we specify the equations of motion for the case of a spinning test particle for a Kerr metric.

According to R.M. Plyatsko \textit{et al.} \cite{plyatsko 1}, the full set of
the exact MPD equations of motion for a spinning test particle in the Kerr
field, if the MP condition (\ref{mp}) is taken into account, obtain a
general scheme for the set of equations of motion for a spinning test particle
in a rotating field. Plyatsko \textit{et al. }\cite{plyatsko 1} \textit{\ }use a set of dimensionless quantities \textit{y}$_{i}$ to achive this. In particular, the Boyer-Lindquist coordinates are represented by
\begin{equation}
y_{1}=\frac{r}{M},\ y_{2}=\theta,\
 y_{3}=\varphi,\ y_{4}=\frac{t}{M}\label{y1}
\end{equation}
the corresponding four-velocity are given by
\begin{equation}
y_{5}=u^{1},\ y_{6}=Mu^{2},\ y_{7}=Mu^{3},\ y_{8}=u^{4}\label{y5}
\end{equation}
and the spatial spin components by \cite{plyatsko 2010}
\begin{equation}
y_{9}=\frac{S_{1}}{mM},\ y_{10}=\frac{S_{2}}{mM^{2}
},\ y_{11}=\frac{S_{3}}{mM^{2}}\label{y9}.
\end{equation}
where $M$ is the mass parameter of the Kerr spacetime. $m$ is the mass of a spinning particle, a constant of motion for the MP SSC and implies that
\begin{equation}
    \frac{dm}{d\tau}=0
\end{equation}

In addition, they introduce the dimensionless quantities
\begin{equation}
x=\frac{s}{M},\ \widehat{E}=\frac{E}{m},\
\widehat{J}=\frac{J_{z}}{mM}.
\end{equation}
representing the proper time $s$ and the constants of motion: Energy ($E$) and the angular
momentum in the $z$ direction ($J_{z}$).

The set of the MPD equations for a spinning particle in the Kerr field is
given by eleven equations. The first four equations are
\begin{equation}
\dot{y}_{1}=y_{5},\ \dot{y}
_{2}=y_{6},\ \dot{y}_{3}=y_{7},\ \dot{y}_{4}=y_{8},
\end{equation}
where the dot denotes the usual derivative with respect to $x$.

The fifth equation is given by contracting the spatial part of equation (\ref{mov1}) with $S_i$ ($\lambda=1,2,3$). The result is multiplied by $S_{1,}S_{2},S_{3}$
and with the MP condition (\ref{mp}) we have the relationships \cite{costa 2015}:
\begin{equation}
 S^{i4}=\frac{u_{k}}{u_{4}}S^{ki} \ \textrm{and} \ S_{i}=\frac{1}{2u_{4}}\sqrt{-g}
\varepsilon_{ikl}S^{kl} \label{Si},  
\end{equation}
then we obtain
\begin{equation}
mS_{i}\frac{Du^{i}}{ds}=-\frac{1}{2}u^{\pi}S^{\rho\sigma}S_{j}R_{\pi\rho
\sigma}^{j}
\end{equation}
which can be written as
\begin{equation}
y_{9}\dot{y}_{5}+y_{10}\dot{y}_{6}+y_{11}
\dot{y}_{7}=A-y_{9}Q_{1}-y_{10}Q_{2}-y_{11}Q_{3},
\end{equation}
where
\begin{equation}
Q_{i}=\Gamma_{\mu\nu}^{i}u^{\mu}u^{\nu},A=\frac{u^{\pi}}
{\sqrt{-g}}u_{4}\epsilon^{i\rho\sigma}S_{i}S_{j}R_{\pi\rho\sigma}^{j}.
\end{equation}
where $g$ is the determinant of the metric $g_{\mu\nu}$.

The sixth equation is given by
\begin{equation}
u_{\nu}\frac{Du^{\nu}}{ds}=0
\end{equation}
which can be written as
\begin{equation}
p_{1}\dot{y}_{5}+p_{2}\dot{y}_{6}+p_{3}
\dot{y}_{7}+p_{4}\dot{y}_{8}=-p_{1}Q_{1}-p_{2}
Q_{2}-p_{3}Q_{3}-p_{4}Q_{4},
\end{equation}
where
\begin{equation}
p_{\alpha}=g_{\alpha\mu}u^{\alpha}.
\end{equation}

The seventh equation is given by
\begin{equation}
E=p_{4}-\frac{1}{2}g_{4\mu,\nu}S^{\mu\nu}
\end{equation}
which can be written as
\begin{equation}
c_{1}\dot{y}_{5}+c_{2}\dot{y}_{6}+c_{3}
\dot{y}_{7}=C-c_{1}Q_{1}-c_{2}Q_{2}-c_{3}Q_{3}+ \widehat{E}
\end{equation}
where

\begin{eqnarray}
c_{1} &=&-dg_{11}g_{22}g_{44}u^{2}S_{3}-d\left(  g_{34}^{2}-g_{33} g_{44}\right)  g_{11}u^{3}S_{2}\nonumber\\
c_{2} &=&dg_{11}g_{22}g_{44}u^{1}S_{3}+d\left(  g_{34}^{2}-g_{33} g_{44}\right)  g_{22}u^{3}S_{1}\nonumber\\
c_{3} &=& d\left(  g_{34}^{2}-g_{33}g_{44}\right)  g_{11}u^{1}S_{2}-d\left(g_{34}^{2}-g_{33}g_{44}\right) g_{22}u^{2}S_{1}\nonumber\\
C &=&g_{44}u^{4}-dg_{44}u^{4}g_{43,2}S_{1}\nonumber\\
&+&d\left(  g_{44}u^{4} g_{43,1}-g_{33}u^{3}g_{44,1}\right)  S_{2} 
+dg_{22}u^{2}g_{44,1}S_{3}
\end{eqnarray}
with
\begin{equation}
d=\frac{1}{\sqrt{-g}}
\end{equation}
where $g$ is the determinant of the metric $g_{\mu\nu}$ and the values of
$g_{11},g_{22},g_{33},...$ are the components of the metric $g_{\mu\nu}$.

The eighth equation is given by
\begin{equation}
J_{z}=-p_{3}+\frac{1}{2}g_{3\mu,\nu}S^{\mu\nu}
\end{equation}
which can be written as
\begin{equation}
d_{1}\dot{y}_{5}+d_{2}\dot{y}_{6}+d_{3}
\dot{y}_{8}=D-d_{1}Q_{1}-d_{2}Q_{2}-d_{3}Q_{4}-\widehat{J}
\end{equation}
where
\begin{eqnarray}
d_{1} &=& -dg_{11}g_{22}g_{34}u^{2}S_{3}+dg_{11}g_{33}g_{34}u^{3}S_{2}\nonumber\\ &+&dg_{11}g_{34}^{2}u^{4}S_{2}-dg_{11}g_{33}g_{44}u^{4}S_{2}\nonumber\\
d_{2} &=& -dg_{11}g_{22}g_{34}u^{1}S_{3}-dg_{22}g_{33}g_{34}u^{3}S_{1}\nonumber\\
&-&dg_{22}g_{34}^{2}u^{4}S_{1}+dg_{22}g_{33}g_{44}u^{4}S_{1}\nonumber\\
d_{3} &=& -dg_{11}g_{34}^{2}u^{1}S_{2}+dg_{22}g_{34}^{2}u^{2}S_{1}\nonumber\\
&+&dg_{22}g_{33}g_{44}u^{2}S_{1}-dg_{11}g_{33}g_{34}u^{1}S_{2}\nonumber\\
D &=& g_{33}u^{3}-dg_{22}u^{2}g_{34,2}S_{1}\nonumber\\
&+&d\left(  g_{44}u^{4}g_{33,1}+g_{11}u^{1}g_{34,1}-g_{33}u^{3}g_{34,1}\right)  S_{2}\nonumber\\
&-& dg_{11}u^{1}g_{34,1}S_{3}.
\end{eqnarray}

Finally, the last three equations are given by%
\begin{equation}
u^{4}\dot{S}_{i}+2\left(\dot {u}_{[4}u_{i]}-u^{\pi
}u_{\rho}\Gamma_{\pi\lbrack4}^{\rho}u_{i]}\right)  S_{k}u^{k}+2S_{n}
\Gamma_{\pi\lbrack4}^{n}u_{i]}u^{\pi}=0
\end{equation}
which give the derivatives of the three spatial components of the spin vector
($\dot{S}_{i}$): $\dot{y}_{9}$, $\dot{y}_{10}$ and $\dot{y}_{11}$. The full set of the exact MPD
equations for the case of a spinning test particle in a Kerr metric under the
Pirani condition (\ref{mp}) is in the appendix of \cite{plyatsko 1}.

After achieving a system of equations of motion for spinning test particles,
we solve them numerically. We use the fourth-order Runge Kutta method for
obtaining the Cartesian coordinates of the trajectories ($x$, $y$,
\thinspace$z$). For our numerical calculations, we take the parameters both of
the central mass and the test particle such as the radio, the energy, the
angular momentum and the components of the four velocity ($u^{\mu}$). We
calculate the full orbit in Cartesian coordinates ($x$, $y$, \thinspace$z$) of
a test particle around a rotating massive body for both spinless and spinning test
particles. Then, we make a comparison of the time that a test particle takes
to do a lap in the two cases.

\subsection{Equations of motion for a spinning test particle orbiting a
massive rotating body}

In the last section, we obtained the general scheme for the set of equations
of motion of a spinning test particle in the gravitational field of a rotating
body \cite{chicone 1}. Now, we consider the case for the equatorial plane, which is given by
the following set of equations:

\begin{equation}
r^{\prime}[s]=\frac{dr}{ds};\ \theta^{\prime}[s]=\frac{d\theta}
{ds}=0;\ \varphi^{\prime}[s]=\frac{d\varphi}{ds};\ 
t\acute{}[s]=\frac{dt}{ds}
\end{equation}
where $s$ is the proper time.

For our numerical calculation, we separate from the full set of equations each one of the
functions for the four velocity vector ($dx^{\mu}/ds$) and the differentials
for the spatial components of spin vector ($S_{i}$). Finally from the eleven differential equations, we obtain the
trajectories of the spinning test particle orbiting around the rotating central mass ($M$). We perform our numerical integration as follow: In the first one, we perform the integration along the direction of the rotation axis of the massive body, and the second one in its opposite sense. The value of the components from initial four
velocity vector is obtained by replacing the values of the constants
of motion ($E$ and $J$), the Carter\'{}s constant ($Q$) and the radio in the Carter\'{}s equations \cite{carter 1}

\begin{equation}
\Sigma \dot{t} =a\left(  J-aE\sin^{2}\theta\right)
\nonumber\\
+ \frac{\left(  r^{2}+a^{2}\right)  \left[  E\left(  r^{2}+a^{2}-aJ\right)
\right]  }{\Delta},\label{car1}    
\end{equation}

\begin{equation}
\Sigma \dot {r}^{2} = \pm R=\pm\left\{
\begin{array}
[c]{c}
\left[  E\left(  r^{2}+a^{2}\right)  \mp aJ\right]  ^{2}\nonumber\\
-\Delta\left[  r^{2}+Q+\left(  J\mp aE\right)  ^{2}\right]
\end{array}
\right\}  ,\label{car2}    
\end{equation}

\begin{equation}
\Sigma \dot{\theta}^{2}  = \pm\Theta=\pm\left\{  Q-\cos^{2}
\theta\left[  a^{2}\left(  1-E^{2}\right)  +\frac{J^{2}}{\sin^{2}\theta
}\right]  \right\}  ,\label{car3}    
\end{equation}

\begin{equation}
 \Sigma \dot{\phi}   =\frac{J}{\sin^{2}\theta}-aE+\frac
{a}{\Delta}\left[  E\left(  r^{2}+a^{2}\right)  -aJ\right]  ,
\label{car4}   
\end{equation}

where $J$, $E$ and $Q$ are constants and

\begin{eqnarray}
\Sigma :=r^{2}+a^{2}\cos^{2}\theta,\nonumber\\
\Delta :=r^{2}+a^{2}-2Mr,\nonumber
\end{eqnarray}
$M$ and $a=J/M$ are the mass and specific angular momentum for the mass unit of
the central source.

The Carter's constant ($Q$) is a conserved quantity of the particle in free fall around a rotating massive body. This quantity affects the latitudinal motion of
the particle and is related to the angular momentum in the $\theta
$\ direction. From (\ref{car3}), one analyzes that in the equatorial plane,
the relation between $Q$ and the motion in $\theta$ is given by

\begin{equation}
\Sigma \dot{\theta}^{2}=Q.
\end{equation}

When $Q=0$ corresponds to equatorial orbits and for the case when $Q\neq0$, one has non-equatorial orbits.

\subsection{MPD equations for a spinless test particle in a Kerr metric}

The traditional form of MP equations is given by the Eq (\ref{mov1}) \cite{mathisson 1} and for our problem, we consider the motion of a spinning test particle in equatorial
circular orbits ($\theta=\pi/2$) of the rotating source. For this case, we
take \cite{plyatsko 2013}
\begin{equation}
u^{1}=0,\ u^{2}=0,\ u^{3}=\textrm{const},\ u^{4}=\textrm{const}
\end{equation}
when the spin is perpendicular to this plane and the MP condition (\ref{mp}),
with
\begin{equation}
S_{1}\equiv S_{r}=0,\ S_{2}\equiv S_{\theta}\neq0,\ 
S_{3}\equiv S_{\varphi}=0.
\end{equation}

The equation is given by
\[
-y_{1}^{3} y_{7}^{2}-2\alpha y_{7}y_{8}+y_{8}^{2}-3\alpha
\varepsilon_{0}y_{7}^{2}+3\varepsilon_{0}y_{7}y_{8}-3\alpha\varepsilon_{0}y_{8}^{2}y_{1}^{-2}
\]
\[
+3\alpha\varepsilon_{0}y_{1}^{2}y_{7}^{4}-\alpha\varepsilon_{0}\left(  1-\frac{2}{y_{1}}\right)  y_{8}^{4}y_{1}^{-3}+\alpha\left(  y_{1}^{6}-3y_{1}^{5}\right)  y_{7}^{3}y_{8}y_{1}^{-3}
\]
\begin{equation}
+\alpha\varepsilon_{0}\left(3y_{1}^{3}-11y_{1}^{2}\right)  y_{7}^{2}y_{8}^{2}y_{1}^{-3}+\varepsilon_{0}\left(-y_{1}^{3}+3y_{1}^{2}\right)
y_{7}y_{8}^{3}y_{1}^{-3}=0\label{c}
\end{equation}

where $y_{1}=r/M$, $y_{7}=Mu^{3}$, $y_{8}=u^{4}$, $\varepsilon_{0}=\left\vert
S_{0}\right\vert /mr$\ and $\alpha=a/M$.

When the particle does not have spin ($\varepsilon_{0}=0$), the set of
equations (\ref{mov1}) with the dimensionless quantities \textit{y}$_{i}$
(\ref{y1}) and (\ref{y5}) is given by
\begin{equation}
-y_{1}^{3} y_{7}^{2}-2\alpha y_{7}y_{8}+y_{8}^{2}=0\label{a}
\end{equation}

In addition to Eq. (\ref{a}), we take the condition $u_{\mu}u^{\mu}=1$ and
obtain
\begin{equation}
-y_{1}^{2} y_{7}^{2}+4\alpha\frac{y_{7}y_{8}}{y_{1}}+\left(  1-\frac{2M}{y_{1}}\right)  y_{8}^{2}=1\label{b}
\end{equation}

We solve the system of equations (\ref{a}) and (\ref{b}) for the case of a
circular orbit and obtain the values of \ $y_{7}=Mu^{3}$ and\ $y_{8}=u^{4}$ in
the equatorial plane.

\section{Gravitomagnetic effects for spinning test particles}

In the study of the gravitomagnetic effects, we find the gravitomagnetic force is the gravitational counterpart to the Lorentz force in electromagnetics. Hence, there is an analogy between classical electromagnetism and general relativity
such as the possibility that the motion of mass could generate the analogous
of a magnetic field.  \cite{costa 2012}. In general relativity, the gravitomagnetic field is caused by mass
current and has interesting physical properties which explain
phenomena such as the precession of gyroscopes or the delay time for test
particles in rotating fields \cite{costa 2008}.

In this section, we describe some phenomena of the trajectories from the spin
vector, represented by a gyroscope, with the help of the gravitomagnetic
effects such as the clock effect, Thomas precession, Lense-Thirring effect or
Sagnac effect \cite{mash 1} \cite{bini 1992} \cite{ciufolini 2013}.

The first effect that we take is the Lense - Thirring effect which has the
consequence that moving matter should somehow drag with itself nearby bo-dies. We can do an analogy of this dragging of mass current with a magnetic field produced by a charge in motion. With this analogy, we set up two spinning test particles orbiting in an equatorial plane of a rotating gravitational field. Then, we compare the trajectories of these two spinning test particles that travel in
opposite directions in the same circular orbit. We found that one of the particles
arrived before the other one. The delay time is due not only to the
dragging of the frame system, but also to the angular motion of the spinning
test particle \cite{faruque 2004}. On the other hand, the rotating
massive body induces rotation and causes the precession of the axis of a
gyroscope which creates a gravitomagnetic field. The form of the figure is the
same, either that the spinning particle orbits in the direction of the central mass
or in opposite direction, but they are out of phase in the space.
\section{Gravitomagnetic clock effect for spinning test particles}
In the scond half of the nineteenth century, Holzm\"{u}ller \cite{holzmuller 1} and Tisserand \cite{tisserand 1} with the help of works in electrodynamics, postuled a gravitomagnetic component for the gravitational influence of the Sun on the motion of planets. The general relativistic effect of the rotation of the Sun with regard to the planetary
orbits was calculated by de Sitter \cite{sitter 1} and later by Lense and Thirring \cite{thirring 1}. After, Ciufolini described the Lense-Thirring
precession of satellites such as LAGEOS and LAGEOS II around the rotating Earth \cite{ciufolini 1}. Then NASA launched a satellite around of the Earth. This satellite was orbiting in the polar plane and carried four gyroscopes whose aim was to measure the drag of inertial systems produced by mass current when the Earth is rotating and to measure the geodesic effect given by curvature of the gravitational field around the Earth \cite{everitt 1}. This experiment was called \textit{Gravity Probe B}.

There is a phenomenon called the gravitomagnetic clock effect which consists
in a difference in the time which is taken by two test particles to travel
around a rotating massive body in opposite directions\ in the equatorial
plane \cite{faruque 2004}. This effect involves the difference in
periods of two test particles moving in opposite directions on the same orbit.
Let $\tau_{+}\left(  \tau_{-}\right)  $ be the proper period that it takes for a test
particle to complete a lap around a rotating mass on a prograde
(retrograde) orbit. In the literature, the majority of works that study the
clock effect consider the difference of periods for spinless test particles.
In this paper, we study the clock effect for two spinning test particles orbiting around to a rotating body in the equatorial plane.

To check our results, we review the papers regarding gravitomagnetic
clock effect \cite{tsoubelis 1} and compare their results with ours.
The delay time given by the clock effect is $t_{+}-t_{-}=4\pi a/c$, where
$a=J/Mc$ is the angular momentum density of the central mass. Tartaglia has studied the
geometrical aspects of this phenomenon \cite{tartaglia 1}, \cite{tartaglia 2001}, and Faruque yields the equation of the gravitomagnetic
clock effect with spinning test particles as \cite{faruque 2004}
\begin{equation}
t_{+}-t_{-}=4\pi a-6\pi S_{0},\label{gm1}
\end{equation}
where $S_{0}$ is the magnitude of the spin.

In true units this relation is given by
\begin{equation}
t_{+}-t_{-}=\frac{4\pi J_{M}}{Mc^{2}}-\frac{6\pi J}{mc^{2}},\label{gm}
\end{equation}
where the first relation of the right could be used to measure $J/M$ directly
for an astronomical body; in the case of the Earth $t_{+}-t_{-}\simeq10^{-7} \textrm {s}
$, while for the Sun $t_{+}-t_{-}\simeq10^{-5}
\textrm {s}
$ \cite{mash 1999}.

\section{Numerical comparison for spinless and spinning test particle via MPD
equations}

We take the set of MPD equations for a spinning test particle in a Kerr me-tric
given in the second section. This set is composed of eleven coupled
di-fferential equations. We input the initial conditions in geometrized units as: $E=0.951906$, $r=10$, $a=1$, $M=1$, Carter\'{}s constant: $Q=0.000008$ and angular momentum: $J=3.426929$. With these initial values, we obtain the four-vector
velocity ($dx^{\mu}/ds$) with the Carter\'{}s equations (\ref{car1} - \ref{car4}). The spatial components of vector spin
($S_{i}$) are used to obtain the integration limits. For our case, the spin components are: $S_1=10^{-12}$, $S_2=0.1$, $S_3=0.1$. Then, we integrate the set of eleven equations, which were presented in the section 2.1, with the
fourth-order Runge-Kutta method \cite{press 1} with a step size of
$\ 2.384185\times10^{-7}$, while Kyrian and Semer\'{a}k integrate with a step
of $M/100=1\times10^{-2}$ \cite{kyrian 1}.\ With this code, we get the
Cartesian coordinates for a circular orbit when the spinning test particle
travels in the same direction of rotation of the central source ($a$) and when it
orbits in opposite sense. We register the coordinate time ($t=x^{4}$) that the
test particle takes to do a lap in each sense of rotation. Finally, we
take the delay time in these two laps and obtained in non-geometrized units
\begin{equation}
\Delta t_{\textrm {spinning}}=t_{+}-t_{-}=2.5212079035\times10^{-8} \textrm {s}.
\end{equation}

Our numerical result is in according with previous works \cite{faruque 2004}, \cite{mash 1}, \cite{bini 2004}, made of an analytic via. In this result, we found that the clock effect is reduced by
the presence of the spin in the test particle. Both Mashhoon \cite{mash 1} and Faruque \cite{faruque 2006} develop an approximation
method for studying the influence of spin on the motion of spinning test
particles \cite{bini 2006}, while we use an integration method of the full set of MPD equations
in order to obtain the value of the coordinate time ($t$).

This delay time is due to the drag of the inertial frames with respect to infinity
and is called the Lense-Thirring effect \cite{mash 1984}. In the case
of the spinning test particles, there is not only a difference in the time
given by the Lense-Thirring effect, but also by a coupling between the angular
momentum of the central body with the spin of the particle \cite{chandra 1}. The features change if the test particle rotates in one
direction or the other; therefore, the period is different for one direction
and for the other, and for whether or not the particle has spin.

\section{Results of the spin vector}

\begin{figure}[ptb]
\centering
\includegraphics[
height=2.6184in,
width=3.0222in
]
{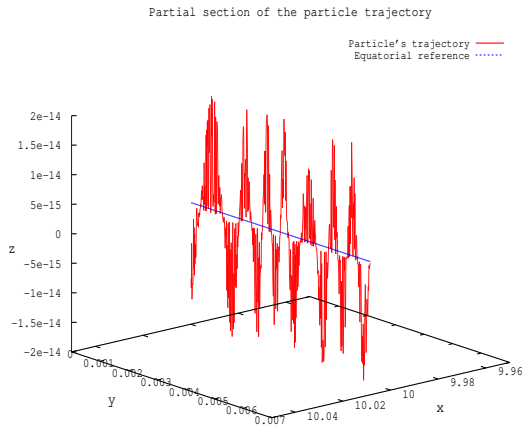}
\caption{The bobbing of the spinning test particle with the Pirani condition}
\label{fig1}
\end{figure}

In regard to the spin tensor ($S^{\mu\nu}$), sometimes, for the numerical
calculation, it is more convenient the spin vector ($S_{i}$) which is given by
the relationship (\ref{Si}). For our
numerical calculations, we take the case when the spin is orthogonal to the
equatorial plane, that is, $S_{1}=0$, $S_{2}\neq0$ and $S_{3}=0$. In this
case, we present our main results with two graphs. For the case ($S_{2}\neq0
$), the spin has a tiny nutation (Figure \ref{fig1}). The first graph shows
the motion of the spin vector in Boyer-Lindquist coordinates ($S_{1},S_{2},S_{3}$).
Since both the radius ($S_{1}$) and the azimuth angle are constant the spin
vector describes an oscillating movement with a maximum height of $2\times
10^{-14}$. This oscillation is very short compared with the radius ($r=10
$) of the circle that describes the trajectory.

\begin{figure}[ptb]
\centering
\includegraphics[
height=2.2344in,
width=3.174in
]
{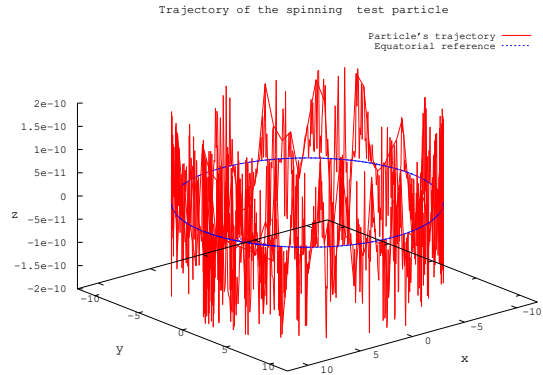}
\caption{The spinning test particle with the Pirani condition has
a helical trajectory}
\label{fig2}
\end{figure}

If we draw at the same time the orbital motion and the spin motion, we obtain
an ascending and descending movement within an enveloping sinusoidal wave
(Figure \ref{fig2}). This movement is called "bobbing" \cite{gralla 1}.
Moreover, this ascendant and descendent movement is due to the supplementary
spin condition that we take to be the MP condition ($S^{\mu\nu}u_{\nu}=0$),
where $u_{\nu}$ is the center of mass four velocity. In this situation, the
center of mass is measured in its proper frame (that is, the frame is at
rest). This phenomenon is due to the shifting of the center of mass, and in
addition, the momentum of the particle not being parallel to its four-velocity in
general. There is a ``hidden momentum" that produces this nutation. In an
analogy with the electric ($E$) and magnetic ($B$) fields, there would be a
$E\times B$ drift, that is, the motion is des-cribed by helical motions
\cite{jackson 1}. Costa \textit{et al.} describe this physical situation due to the MP supplementary condition \cite{costa 1}.
In the case that we are studying, the world tube is formed by all possible
centroids which are determined by the MP spin supplementary condition. The size
of this tube is the minimum size of a classical spinning particle without
violating the laws of Special Relativity. Additionally, this world tube
contains all the helical solutions within a radius $R=S/M$.
The electromagnetic analogue of the hidden momentum is $\overrightarrow{\mu
}\times\overrightarrow{E}$ which describes the bobbing of a magnetic dipole
orbiting a cylindrical charge. Let the line charge be along the $z$ axis, the
$\overrightarrow{E}$ the electric field it and a charged test particle with
magnetic dipole moment $\overrightarrow{\mu}$ orbiting it. The particle will
have a hidden momentum $\overrightarrow{P}_{hid}=$ $\overrightarrow{\mu}%
\times\overrightarrow{E}$ and oscillates between positive and negative values
along the $z$ axis with ascendant and descendent movements in order to keep
the total momentum constant.
According with other papers that use the MPD equations, the novelty of our
work is that we calculate numerically the full set of MPD equations for the
case of a spinning test particle in a Kerr metric. Secondly, we take the spin
without restrictions in its velocity and spin orientation. In the paper by
Kyrian and Semer\'{a}k the third example that they work, refers to the
particular case when the spin is orthogonal to the equatorial plane in a Kerr
metric \cite{kyrian 1}.
On the other hand, our interest it is not only in to describe the trajectories of
spinning test particles, but also to study the gravitomagnetic clock effect
via the MPD equations. Therefore, we calculate numerically both trajectories,
i.e., in the same and in the opposite way, in the case of a circular orbit. We
measure the delay time for three different situations, namely, for the
spinless test particle, traveling one way and the opposite, and for the
spinning test particle for two different spin configuration. Regarding the
spin configuration for the first, the spin has its value orthogonal to the
equatorial plane of the trajectory, and for the second, the spin has no restrictions.
For the numerical solution, we give the full set of MPD equations explicitly,
while that Kyrian and Semerak only name them. Also, we give the complete
numerical solution. In the majority of cases, the solutions are partial
because it is impossible to solve analytically a set of eleven coupled
differential equations. The recurrent case that they solve is a spinning test
particle in the equatorial plane and its spin value is constant in the time
($S_{\perp}=$ constant) \cite{mash 1}.

\section{Conclusions}

In this paper, we take the Mathisson-Papapetrou-Dixon (MPD) equations yielded
by Plyatsko \textit{et al.} and apply them to the case of a spinning test
particle orbiting around a rotating massive body in an equatorial plane.
In addition, we yield a scheme for the eleven equations of the full set of
equations of motion when the particle is orbiting any gravitational
field. In the second part, we calculate the numerical solution of the
trajectories in Cartesian coordinates ($x$, $y$, \thinspace$z$) of the
spinning test particles orbiting in a Kerr metric and compare the time of two
circular orbits in the equatorial plane for two test particles that travel in
the same orbit but in opposite directions. There is a delay time for a fixed
observer relative to the distant stars. This phenomenon is called clock
effect. For the case of the spinning test particles, this delay time is given
not only by the angular momentum from the central mass, but also by the couple
between the angular momentum from the massive rotating body and the parallel
component of the spin of the test particle. In the MPD equations, this couple
is given by the contraction between the components of the Riemman tensor
($R^{\mu}$ $_{\nu\rho\sigma}$) and the spin tensor ($S^{\rho\sigma}$).

On the other hand, we obtained the graphs that describe both the orbital
motion and the motion of the spin vector freely rotating in the polar
component ($S_{2}\neq0$). With this kind of motion, we make an analogy with
the bobbing of a magnetic dipole in a electromagnetic field.

In the future, we will work on the set of equations of motion of a test
particle both spinless and spinning for spherical orbits, that is, with
constant radius and in non-equatorial planes in a Kerr metric. In addition, we
are interested in relating these equations with the Michelson - Morley type experiments.

\end{document}